\documentclass[runningheads]{llncs}
\usepackage[T1]{fontenc}
\usepackage{graphicx}
\usepackage[utf8]{inputenc}
\usepackage{amsmath, amssymb }
\usepackage{url}
\usepackage{cite}
\usepackage{color}
\usepackage{xcolor}
\usepackage{multirow}
\usepackage{booktabs}
\usepackage{siunitx}
\usepackage{microtype}

\begin{document}
\title{AnalysisGNN: Unified Music Analysis \\ with Graph Neural Networks}
%
%
\author{Emmanouil Karystinaios\inst{1}\thanks{* Equal contribution.}, Johannes Hentschel\inst{2}$^*$, \\Markus Neuwirth\inst{2}, Gerhard Widmer\inst{1}}
\authorrunning{E. Karystinaios et al.}
%
\institute{Institute of Computational Perception, Johannes Kepler University Linz, Austria \\
\email{firstname.lastname@jku.at}\\
\and Anton Brukner University Linz, Austria\\
\email{firstname.lastname@bruckneruni.at}\\
}
\maketitle            

\begin{abstract}
Recent years have seen a boom in computational approaches to music analysis, yet each one is typically tailored to a specific analytical domain. In this work, we introduce \emph{AnalysisGNN}, a novel graph neural network framework that leverages a data‐shuffling strategy with a custom weighted multi‐task loss and logit fusion between task‐specific classifiers to integrate heterogeneously annotated symbolic datasets for comprehensive score analysis. We further integrate a Non‐Chord‐Tone prediction module, which identifies and excludes passing and non‐functional notes from all tasks, improving the consistency of label signals. Experimental evaluations demonstrate that AnalysisGNN achieves performance comparable to traditional static‐dataset approaches, while showing increased resilience to domain shifts and annotation inconsistencies across multiple heterogeneous corpora. The full source code for this work is available at: \url{github.com/manoskary/analysisgnn}

\end{abstract}

\section{Introduction}

Symbolic music analysis is a cornerstone of music information retrieval and musicology. Traditional methods for harmonic analysis, cadence detection, and phrase segmentation often rely on rule‐based systems or statistical models. While recent deep learning approaches have improved performance by leveraging relatively large annotated datasets or self‐supervised pretraining, they typically treat each analysis problem separately, thereby missing the inherent interdependencies in musical structure.

Multi‐task learning offers a promising direction to exploit cross‐task knowledge transfer—for example, features learned during harmonic analysis may enhance cadence detection—but in practice, available datasets focus on single musical elements (harmony, cadence, or hierarchical structure) and suffer from inconsistent annotations. To address these challenges, we replace task‐incremental pipelines with a unified data‐shuffling strategy: during training, mini‐batches are sampled across all tasks, with a custom weighted multi‐task loss that balances each objective, and logits from task‐specific classifiers are fused to reinforce shared representations. Additionally, we incorporate a Non‐Chord‐Tone prediction branch that filters out passing and non‐functional notes before downstream tasks, yielding cleaner input signals and reducing conflicting labels.

Graph Neural Networks (GNNs) have emerged as a powerful solution for modeling the complex, non‐sequential relationships inherent in music scores. For example, \emph{ChordGNN}~\cite{Karystinaios2023_RomanNumeralAnalysis} represents notes as nodes in a graph to capture Roman numeral harmony, achieving competitive results compared to previous approaches. Similarly, graph‐based methods have improved cadence detection by directly leveraging musical context.

Our proposed framework, \emph{AnalysisGNN}, builds on these ideas by combining a data‐shuffling training regimen with weighted multi‐task optimization and inter‐classifier logit fusion, alongside a non-functional note exclusion mechanism. This design addresses the limitations of fragmented, specialized datasets and isolated models through a single, cohesive Graph Neural Network. The resulting system achieves good performance on each analysis task, including cadence detection, Roman numeral analysis, section and phrase identification, and metrical position estimation, while also maintaining robustness against domain shifts and annotation inconsistencies.

Our contributions are four‐fold:
\begin{enumerate}
  \item We propose a training strategy and architecture for heterogeneous multi‐task symbolic music analysis..
  \item We introduce a number of new music analysis tasks such as Non‐Chord‐Tone prediction module that identifies and filters passing and non‐functional notes.
  \item We assemble and preprocess the largest compilation of heterogeneously annotated symbolic music datasets, unified into a graph representation suitable for GNN processing.
  \item We demonstrate that AnalysisGNN achieves competitive performance while exhibiting strong resilience to domain shifts and annotation variability.
\end{enumerate}


\section{Related Work}\label{sec:related}

\subsection{Harmonic Analysis (Roman Numeral Labeling)}
Automatic analysis of functional harmony has long been studied in the fields of music information retrieval and music theory. 

Early systems employed rule-based grammars, probabilistic models, dynamic programming, and grammar induction for harmonic and metrical analysis of music~ \cite{Temperley2001_CognitionBasicMusical,Temperley2009_UnifiedProbabilisticModel,Raphael2004_FunctionalHarmonicAnalysis}.
The advent of deep learning brought significant improvements. Micchi et al. \cite{MicchiGG20_PitchRepresentation_TISMIR} trained one of the first neural networks (a convolutional recurrent model) for Roman numeral analysis, and Nápoles López et al.  
\cite{NapolesLopez2021_AugmentedNetRomanNumeral} later introduced \emph{AugmentedNet}, a CNN-LSTM architecture that enhanced performance via data augmentation and multi-task learning. Subsequently, two more systems using multitask approaches for Roman numeral analysis were introduced using techniques to mitigate the interdependency between the Roman numeral prediction subtasks~\cite{micchi2021deep, mcleod2021modular}.

Despite these advances, purely sequential models struggle with the inherent complexity of polyphonic scores---they require the score to be serialized or binned into time slices, which risks losing voice-leading details and long-range dependencies. 
This limitation has motivated the use of graph neural networks.  
\emph{ChordGNN}~\cite{Karystinaios2023_RomanNumeralAnalysis} constructs a graph of all notes in a piece and applies graph convolution to aggregate note features into chord predictions.  
It treats Roman numeral analysis as a multi-task classification problem (predicting multiple components of the Roman numeral label) similar to previous approaches and introduces an edge contraction algorithm to pool information from note-level to chord-level representations.  
As a result, \emph{ChordGNN} achieved high performance on Roman numeral analysis, outperforming previous CNN/CRNN models on standard datasets. The most recent approach by Sailor~\cite{sailorrnbert} has used a transformer model and BERT-like pretraining to build a more robust encoder, which has in turn resulted in the current state-of-the-art performance.

\subsection{Cadence Detection and Phrase Segmentation}
Cadences serve as musical punctuation, marking phrase or sectional boundaries.  
Traditional cadence detection methods relied on hand-crafted features and classifiers, such as SVMs trained on features representing intervallic sequences that form full (or authentic) or half cadences ~\cite{bigo2018relevance}.  
More recent approaches, such as those by Karystinaios and Widmer \cite{Karystinaios2022_CadenceDetectionSymbolic}, reformulated cadence detection as a node classification task on a note-level graph representation of the score.  
In their work, a GNN was trained to predict cadence likelihood for each note (or group of notes), capturing non-local context through message-passing.  
Derivative work has further refined the process by introducing more sophisticated training techniques and models that have resulted in better performance~\cite{Karystinaios2024_GraphMuseLibrarySymbolic}, although the lack of a unified benchmark prevents a direct comparison.

\subsection{Graph Neural Networks in Music Information Retrieval}

Beyond analysis, GNNs have been applied in other MIR tasks, such as music recommendation, emotion modelling, music generation, and performance modeling.  
For example, Jeong et al. \cite{Jeong2019_VirtuosoNetHierarchicalRNNbased} combined a GNN with a hierarchical RNN to model expressive piano performance, capturing both local note interactions and broader musical context.  
Similarly, graph-based representations have been exploited in music generation, where hierarchical graphs of chord progressions and rhythmic patterns help maintain long-term coherence in generated compositions~\cite{lim2024hierarchical, cosenza2023graph}. 
These developments underscore the flexibility of graph-based approaches and their potential to unify disparate music processing tasks under a single framework.




\subsection{Unified Multi-Task Learning Frameworks}\label{subsec:cl_rel}

Multi-task learning (MTL) trains a single model on multiple related objectives simultaneously, leveraging shared representations to improve performance and generalization across tasks~\cite{zhang2021survey}. By interleaving training examples from heterogeneous datasets, rather than processing tasks sequentially, unified MTL avoids the catastrophic forgetting endemic to task-incremental pipelines and mitigates domain shifts between specialized corpora.

A central challenge in unified MTL is balancing conflicting gradients and annotation schemes. Common solutions include: i) tailor optimization by assigning task-specific weights to losses or gradients, either heuristically or via uncertainty estimation, to modulate each objective’s contribution to the total loss~\cite{liu2023famo, liebel2018auxiliary}; ii) dynamic task sampling by shuffling mini-batches across tasks according to dataset size, task difficulty, or performance criteria, ensuring stable gradient updates and preventing domination by any single task~\cite{guo2018dynamic}, iii) logit-level fusion by combining the raw outputs of task-specific classifiers to reinforce shared features and encourage co-adaptation between tasks~\cite{mishra2021cross}.

In symbolic music analysis, auxiliary prediction branches can further improve consistency by filtering noisy labels. For example, a Non-Chord-Tone (NCT) detection head can identify and exclude passing or non-functional notes before they propagate to downstream tasks, reducing label conflicts and sharpening the signal for harmony, cadence, and phrase analyses. Graph Neural Network architectures such as \emph{ChordGNN}~\cite{Karystinaios2023_RomanNumeralAnalysis} have demonstrated the efficacy of combining shared graph encoders with modular task heads; our framework extends this paradigm with data-shuffling, custom weighting, logit fusion, and NCT filtering to achieve robust, unified score analysis.

\section{Methodology}\label{sec:methodology}

\subsection{Model Architecture}

\begin{figure}
    \centering
    \includegraphics[width=\linewidth]{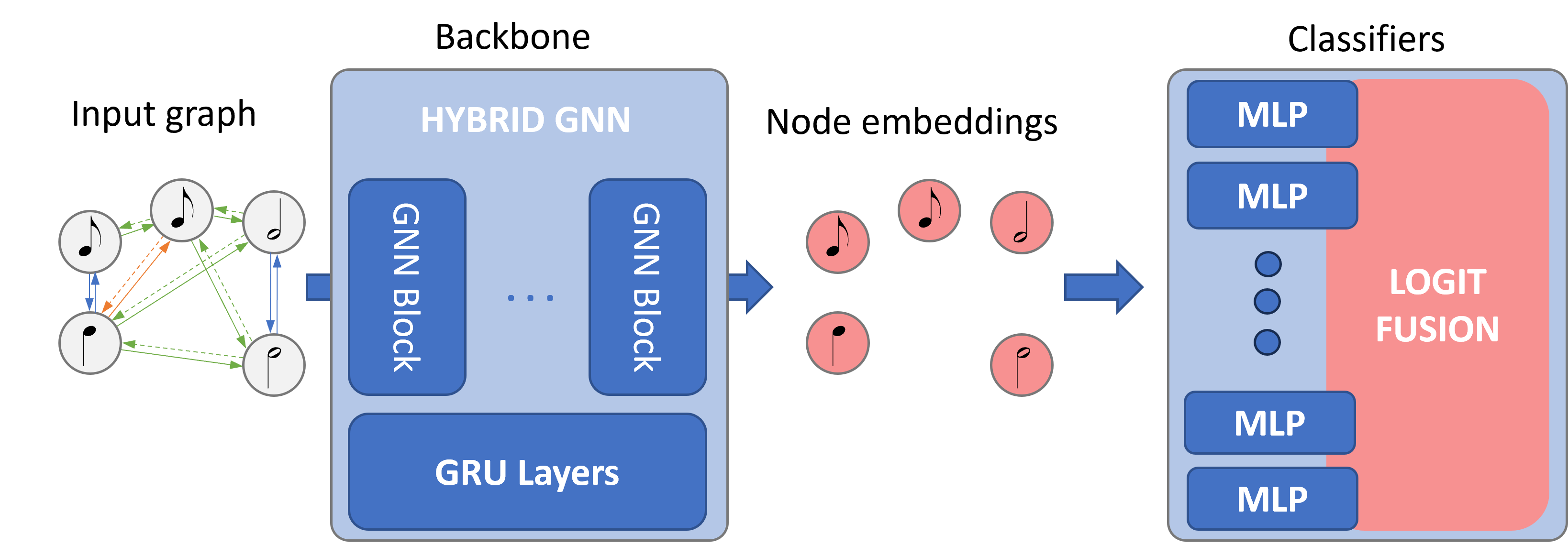}
    \caption{Architecture overview: the input score graph is processed by stacked Heterogeneous GCN blocks in parallel with note-sequence features through an equally deep GRU; their shared embedding feeds 2-layer MLP task heads, whose outputs are then refined via a cross-task logit-fusion layer.
    }
    \label{fig:model-arch}
\end{figure}

For our model, we leverage previous graph-based architectures from the literature used for music analysis while integrating our contributions of logit fusion to architecture design. Input to our model is music scores represented as graphs, where vertices are notes and edges represent temporal relations between those notes, similar to ~\cite{Jeong2019_VirtuosoNetHierarchicalRNNbased, Karystinaios2022_CadenceDetectionSymbolic}.
The backbone of our model is a Hybrid Graph Neural Network similar to the one introduced in \cite{Karystinaios2024_GraphMuseLibrarySymbolic}. In more detail, the score graph enters a series
of Heterogeneous Graph Convolutional blocks. In parallel, the note features are segmented
per piece and padded where necessary and then they are passed through a sequential model such as a GRU with the same number of layers as the number of graph convolutions. The common representation
is then used by a series of 2-layer MLP classifiers. Finally, a logit-fusion layer is added
where the logit prediction of each task communicate with each other. A sketch of the architecture is shown in Figure~\ref{fig:model-arch}.

The HybridGNN encoder combines a sequential model (GRU) with a Graph Convolutional Network (GCN) to effectively capture the hierarchical structure inherent in musical scores. In this design, the GCN integrates information at multiple levels—notes, beats, and measures—thereby functioning as a heterogeneous graph neural network. For note-type nodes, the input features are distilled to essential attributes such as pitch spelling and duration. In contrast, the features for beats and measures are computed as the mean values of the features from their associated notes. Despite this layered encoding, all predictions are ultimately made at the note level.

\subsection{Training}\label{subsec:training}

Training \emph{AnalysisGNN} proceeds by interleaving examples from all tasks via a data-shuffling strategy, combined with a custom weighted multi-task loss and logit-level fusion. At each update step, mini-batches are sampled across the current set of tasks $\mathcal{T}$, ensuring that no single task dominates the optimization.  

We adopt a dynamically weighted cross-entropy loss similar to~\cite{liebel2018auxiliary}, but normalized by the number of tasks:
\begin{equation}
\mathcal{L}_{\mathrm{clf}}
\;=\;
\frac{1}{|\mathcal{T}|}
\sum_{t \in \mathcal{T}}
\biggl(
  \frac{\mathcal{L}_t}{2\,\sigma_t^2}
  \;+\;
  \log\bigl(1 + \sigma_t^2\bigr)
\biggr),
\end{equation}
where $\mathcal{L}_t$ is the cross-entropy for task $t$ and each $\sigma_t$ is a learnable scale controlling the task’s weight.  This formulation both balances gradients from heterogeneous annotation schemas and regularizes the learned weights to prevent any $\sigma_t$ from collapsing to zero.

To encourage shared representations across tasks, we apply a logit‐level fusion mechanism that refines each task’s raw outputs by integrating information from all other task heads.  Concretely, for each task \(t\) in the current set \(\mathcal{T}\):
\[
\begin{aligned}
\mathbf{z}_t &= \mathrm{Clf}_t(\mathbf{h}), 
&&\text{(raw logits)}\\
\mathbf{p}_t &= \mathrm{Proj}_t(\mathbf{z}_t),
&&\text{(projection into a common \(d\)-dimensional space)}
\end{aligned}
\]
We then stack all projections into a matrix
\[
P = 
\begin{bmatrix}
\mathbf{p}_1 \\[3pt]
\mathbf{p}_2 \\[-2pt]
\vdots \\[2pt]
\mathbf{p}_{|\mathcal{T}|}
\end{bmatrix}
\in\mathbb{R}^{|\mathcal{T}|\times d}
\]
and refine them via a transformer‐style self‐attention layer:
\[
\widetilde{P}
=\mathrm{LayerNorm}\Bigl(P \;+\;\mathrm{MultiHeadAttn}(P,\,P,\,P)\Bigr),
\]
where \(\mathrm{MultiHeadAttn}\) denotes standard multi‐head attention and the LayerNorm adds a residual connection.

Finally, each task’s refined logits are obtained by selecting its row from \(\widetilde P\) and passing it through the corresponding fusion head:
\[
\hat{\mathbf{z}}_t \;=\; \mathrm{Fusion}_t\bigl(\widetilde{P}_t\bigr),
\]
where \(\widetilde{P}_t\) is the \(t\)-th row of \(\widetilde{P}\).  We then compute the cross‐entropy loss on \(\hat{\mathbf{z}}_t\) for each task \(t\). This attention‐based fusion allows each task head to attend to and integrate information from all other tasks by sharing confidence and contextual cues, leading to more coherent multi‐task representations.  

In addition to the main analysis tasks, we introduce a Non-Chord-Tone (NCT) prediction branch. This auxiliary head labels each note as functional or non-function in regards to the underlying harmony and structure, with its own cross-entropy loss $\mathcal{L}_{\mathrm{NCT}}$.During training, we predict NCT labels but do not mask out non‐chord‐tones in the multi‐task losses, since this preserves passing‐note examples in the gradient signal and prevents the model from collapsing (by masking everything) or misclassifying functional notes.  

At inference time, however, we can leverage the NCT predictions as a gating mechanism by  classifying each note as chord‐tone or non‐chord‐tone, and pass only those identified as chord‐tones to the task‐specific heads. This selective inference reduces computational overhead and mitigates error propagation by focusing predictions on musically functional notes.

\section{Corpora}\label{sec:datasets}

\subsection{Overview and Preprocessing}

\begin{table*}[htb]
\resizebox{\textwidth}{!}{%
\begin{tabular}{@{} l |
                   S[table-format=4.0] 
                   S[table-format=7.0] 
                   S[table-format=6.0] 
                   S[table-format=6.0] 
                   |
                   S[table-format=3.0] 
                   S[table-format=6.0] 
                   S[table-format=5.0] 
                   |
                   S[table-format=4.0] 
                   S[table-format=7.0] 
                   S[table-format=6.0] 
                   @{}}
\toprule
\multicolumn{1}{r|}{dataset} & \multicolumn{4}{l|}{\textbf{Distant Listening Corpus (DLC)}} & \multicolumn{3}{l|}{\textbf{AugmentedNet}} & \multicolumn{3}{l}{\textbf{Combined Cadence Datasets}} \\
task & {pieces} & {notes} & {labels} & {labels*} & {pieces} & {notes} & {labels} & {pieces} & {notes} & {labels} \\ \midrule
\textbf{note level} & 1266 & 2060662 &  &  & 353 & 758555 &  & 100 & 131418 &  \\
\textbf{chord} & 1266 & 2034151 & 318781 & 234667 & 353 & 757989 & 88559 & 0 & 0 & 0 \\
\textbf{phrase} & 1265 & 2060348 & 21395 & 15575 & 0 & 0 & 0 & 0 & 0 & 0 \\
\textbf{cadence} & 916 & 1334356 & 13908 & 9540 & 0 & 0 & 0 & 100 & 131418 & 4147 \\
\textbf{pedal} & 1266 & 2034151 & 3436 & 2632 & 0 & 0 & 0 & 0 & 0 & 0 \\ \bottomrule
\end{tabular}
}
\caption{\textbf{Dimensions of the three datasets.} 
The note-level row corresponds to the total number of graphs (pieces) and graph nodes (notes). This row does not report labels because the target labels on the note level either derive from the score (total number of notes) or from the harmony annotations (number of notes in the chord row).
The note numbers in the rows below reflect for how many graph nodes a valid label was present for the corresponding task.
The \textit{labels} subcolumns correspond to the number of labels that have been included in each dataset's graphs.
\textit{labels*} is present only for the DLC for which repeats have been expanded; this column indicates the number of labels present in the original files before the expansion.
}
\label{tab:datasets}
\end{table*}

Our work combines multiple datasets with diverse analytical annotations to train a single model: the AugmentedNet dataset~\cite{NapolesLopez2021_AugmentedNetRomanNumeral}, 
the Distant Listening Corpus (DLC)~\cite{Hentschel2025_DistantListeningCorpus}, the Bach WTC cadence dataset~\cite{giraud2015computational}, the Mozart string quartet cadence dataset~\cite{allegraud2019learning}, and the Haydn string quartet cadence dataset~\cite{sears2018simulating}.
The DLC provides 1266 symbolic music scores in MuseScore format, featuring internally consistent annotations (i.e. chords, phrases, and cadences) integrated directly within the scores.
AugmentedNet contributes 353 pieces, offering Roman numeral harmonic analyses in RomanText \texttt{rntxt} format linked to scores parsed from various formats, representing an aggregation of multiple analysis sources.
The remaining datasets feature only cadence annotations which have been inserted into musicXML scores for ease of processing. For simplicity, hereforth we refer to the union of the three datasets that contain solely cadence annotations as the cadence dataset.
In order to facilitate the compilation of a unified graph dataset, the note and annotation content of all datasets was processed into a shared tabular representation.

Since the cadence dataset has no overlap with the DLC and AugmentedNet dataset, we focus on the processing of the two latter. Overlapping pieces between the two datasets were identified, and pieces present in the predefined AugmentedNet test set were excluded from our training data derived from the DLC to allow for later comparison.
Key analytical information, such as Roman numerals (simplified to root and quality), local keys, and chord inversions, was unified across datasets into common formats.
Additional features relevant for analysis, like note-level scale degrees and downbeat information, were derived for both datasets.

The final combined dataset comprises the processed versions of all included pieces from all sources, retaining the distinct annotations for pieces present in all datasets (excluding the test set conflicts). This results in a total of 1719 annotated pieces available for our study (cf. Table~\ref{tab:datasets}).

\subsection{Handling Data}

For all our datasets—including Cadence (which comprises several subdatasets), DCL, and AugNET—the tabular data (described in the previous section) is first converted into graph representations. In these graphs, each node corresponds to a musical note, and relevant analytical labels are propagated to the nodes using the GraphMuse Python package~\cite{Karystinaios2024_GraphMuseLibrarySymbolic}. Furthermore, we apply transposition augmentation to our training data while preserving pitch spelling sensitivity, resulting in an approximate tenfold increase in available scores.

During training, the model is exposed to the entire collection of datasets for a fixed number of iterations, with every batch containing a combination of all three datasets and all tasks. When it comes to
validation and testing, we assign one test and validation split per dataset. Accordingly, each dataset is divided into training, validation, and test sets, with the test set covering
roughly $20\%$ of the total data. The AugmentedNet dataset provided predefined data splits for training, validation, and testing, while for other two datasets we perform a random split.

As detailed in Table~\ref{tab:datasets}, each dataset presents a unique set of annotations and tasks. Certain datasets (particularly DLC) include pieces with partial, missing, or invalid annotations. To manage these issues, we frame each score graph as a semi-supervised node classification problem, masking out any invalid or missing labels at the note level. For example, in the DLC dataset, some pieces lack cadence annotations and certain notes may have missing or unencodable harmony labels. By masking these problematic samples during training, we prevent invalid data points from influencing the model while still leveraging all available notes for predicting the valid labels.

\subsection{Tasks}

In our framework, we address a wide range of tasks that include both tasks already available in the literature and new tasks introduced in this work. Although our model makes predictions at the note level—assigning a label for each task to every note—the scope of these annotations varies: some apply uniformly to all notes sharing the same onset, others apply to individual notes, and some extend over longer segments of the score.

Specifically, our GNN-based model predicts 20 distinct properties for each note (or graph node). However, the prediction can also be computed at the onset or beat level on demand. The predictions include detecting the presence and type of cadences, identifying phrase and section boundaries, flagging pedal points, assessing metrical strength, and marking harmony onsets/changes. In addition, we predict harmonic analysis features as defined in~\cite{NapolesLopez2021_AugmentedNetRomanNumeral}, such as local key, tonicization key, root, bass, harmonic rhythm, inversion, quality, pitch-class set, common Roman numeral, and chord degree.

Beyond these established tasks, we introduce novel note-level tasks aimed at capturing the functional role of each note within the underlying harmony. For instance, our framework determines boolean properties indicating whether a note functions as the bass, the root, or is part of the expected chordal structure. For example, in a "C:I" annotation, the note C is identified as the root, whereas a passing note D is not a constituent of the chord—even though both may be consistently labeled as "C:I" by the model. By leveraging the granularity afforded by our GNN, we enrich the annotations with musicologically informed properties that offer deeper insight into the model’s understanding of harmonic structure.

\section{Experiments}\label{sec:experiments}

\begin{table*}[bt]
\centering
\resizebox{\textwidth}{!}{
\begin{tabular}{c|l|c|c|c|c|c|c}
\textbf{Dataset} & \textbf{Model} & \textbf{Cadence} & \textbf{Roman} & \textbf{Phrase} & \textbf{Pedal} & \textbf{Metr.} & \textbf{Section} \\
\hline
\multirow{3}{*}{Cadence} 
    & GraphMuse               & \textbf{.516} & --  & --  & --  & --  & -- \\     
    & AnalysisGNN (multi-corpus)   & .497         & --  & --  & --  & --  & -- \\
\hline\hline
\multirow{4}{*}{AugNet} 
    & AugmentedNet            & --    & .464 & --  & --  & --  & -- \\ 
    & ChordGNN+Post           & --    & .518 & --  & --  & --  & -- \\ 
    & RNBert                  & --    & \textbf{.574} & --  & --  & --  & -- \\     
    & AnalysisGNN & --    & .530 & --  & --  & --  & -- \\ 
\hline\hline
\multirow{2}{*}{DLC} 
    & RNBert                  & --    & .301 & --  & --  & --  & -- \\     
    & AnalysisGNN  & .558 & \textbf{.516} & .742 & .771 & .761 & .768 \\     
\end{tabular}
}
\caption{Overview of Models by Target Task. For each dataset, we compare baseline models—trained and evaluated on their respective datasets—with AnalysisGNN, which is trained on all three datasets. Cadence, phrase, pedal point, and section are evaluated using note-level macro F1 score; Roman numeral predictions are assessed with the CSR score~\cite{NapolesLopez2021_AugmentedNetRomanNumeral}; and metrical strength is measured by accuracy. A Roman numeral is considered correct only when its local key, degree, quality, and inversion are all predicted accurately.}
\label{tab:model_overview}
\end{table*}

\subsection{Comparison to Single-Corpus Approaches}

\begin{table}[bt]
\centering
\begin{tabular}{l|c|c|ccc}
\toprule
\multirow{2}{*}{\textbf{Trained on}} 
  & \multicolumn{1}{c|}{\textbf{Eval on Cadence}} 
  & \multicolumn{1}{c|}{\textbf{Eval on AugNet}} 
  & \multicolumn{3}{c}{\textbf{Eval on DLC}} \\
  & Cad. F1 & RN & Cad. F1 & Phrase  & RN \\
\midrule
Cadence only           & \textbf{.516} & – & – & – & – \\
AugNet only            & - & .515 & – & – & .441 \\
DLC only               & .479 & .503 & .556 & \textbf{.751} & \textbf{.563} \\
All corpora (combined) & .497 & \textbf{.530} & \textbf{.558} & .742 & .516 \\
\bottomrule
\end{tabular}
\caption{Cross‐corpus evaluation of AnalysisGNN. Rows indicate the training set (single‐corpus or combined) and columns report performance on each target corpus, broken down by task.}
\label{tab:cross_corpus}
\end{table}

To assess the benefits of our unified analysis framework, we compare AnalysisGNN against models trained on individual corpora. In our experiments, we train AnalysisGNN on the combination of all available datasets. Table~\ref{tab:model_overview} presents a comparative evaluation of AnalysisGNN alongside established baselines.

For the AugmentedNet dataset, AnalysisGNN is benchmarked against state-of-the-art models from the literature, including AugmentedNet~\cite{NapolesLopez2021_AugmentedNetRomanNumeral}, ChordGNN~\cite{Karystinaios2023_RomanNumeralAnalysis}, and RNBert~\cite{sailorrnbert}. In the context of cadence detection, we compare against the HybridGNN variant from GraphMuse~\cite{Karystinaios2024_GraphMuseLibrarySymbolic}. Notably, for the DLC dataset, characterized by heterogeneous annotations compared to the AugmentedNet, AnalysisGNN is uniquely capable of effectively handling and comparing the diverse labels. We observe that although AnalysisGNN does not perform as well with single-corpus models, it demonstrates robust, mean performance across tasks, underscoring the advantage of a unified approach in mitigating domain shifts and annotation discrepancies. 

We underline the effect of the domain shift of datasets by showcasing the drop in performance of RNBert when evaluated on the Roman numeral test-set of the DLC corpus. To investigate further whether AnalysisGNN is robust to domain shifts or whether there is any positive knowledge transfer between inter-dataset tasks we present in Figure~\ref{tab:cross_corpus} single corpora vs joint corpora AnalysisGNN models and their performance. Although the DLC dataset contains many tasks, we only showcase the common tasks between datasets such as Cadence and Roman numeral prediction. In all single corpus models, we observe the effect of the data domain shift, where performance drops significantly when the model is evaluated for a learned task on a different dataset. However, the joint-corpus model resists to such domain shift only showcasing minor performance drop for Roman numeral prediction and even positive knowledge transfer on the Cadence task for the DLC test set.

\subsection{Configuration Study}

In order to determine the impact of various components and training strategies for AnalysisGNN, we conducted an extensive configuration study. Our experiments focused on selecting optimal hyperparameters as well as evaluating different model variants and ablation settings. Table~\ref{tab:configuration_study} provides a summary of the configurations tested along with their performance in terms of Roman numeral prediction, Cadence, and Phrase on both the DLC and AugmentedNet corpora. 

\begin{table}[!b]
    \centering
    \resizebox{\textwidth}{!}{
    \begin{tabular}{r||c|c|c|c}
       \textbf{Configuration}  & \textbf{RN (DLC)} & \textbf{RN (AugNET)} & \textbf{Cadence (DLC)} & \textbf{Phrase (DLC)}\\
       \hline
         w/o Logit Fusion   & .503 & .491 & .541 & \textbf{.752}\\
         w/o Transpositions  & .416 & .366 & .418 & .687\\
         w/o Aux-Tasks      & .506 & .511 & .532 & .723\\         
         \hline
         full AnalysisGNN   & \textbf{.516} & \textbf{.530} & \textbf{.558} & .742
    \end{tabular}}
    \caption{Configuration study for AnalysisGNN. RN (DLC) and RN (AugNET) show the CSR accuracy of Roman Numeral prediction on the DLC and AugmentedNet corpora respectively. We also report Cadence and Phrase macro F1 scores on the DLC test set.}
    \label{tab:configuration_study}
\end{table}

Overall, the ablation results highlight several key insights. Removing the logit fusion layer leads to a modest drop in Roman numeral and cadence performance, indicating that cross-task attention helps reconcile conflicting gradients and share contextual cues, even though phrase detection can sometimes benefit from more task-specific signals. 
Removing transposition augmentation produces the largest overall decline, highlighting the importance of pitch invariance, a result consistent with AugmentedNet’s findings~\cite{NapolesLopez2021_AugmentedNetRomanNumeral}.  

Removing auxiliary tasks, we notice an overall performance drop which indicates that tasks induce positive transfer among them. For example knowing which notes are structurally relevant sharpens harmonic analysis, and conversely, learning harmonic context helps the model distinguish chord‐tones from embellishments. These cross‐task gains go beyond harmonic analysis to structural elements such as cadences, phrases and sections.

\subsection{Configuration}

All our models are trained on a single Nvidia RTX A6000 GPU with the hierarchical neighbor sampling strategy introduced in ~\cite{Karystinaios2024_GraphMuseLibrarySymbolic} with a subgraph size of 500 and a batch size of 250. The encoder is a 3-layer HybridGNN with a hidden size of 256 and an output size of 128. We trained with AdamW optimizer, weight decay of 0.0005 and a learning rate of 0.005 with linear warmup of 500 steps and then switched to cosine annealing scheduler. 




\begin{figure}[t]
    \centering
    \includegraphics[width=\linewidth]{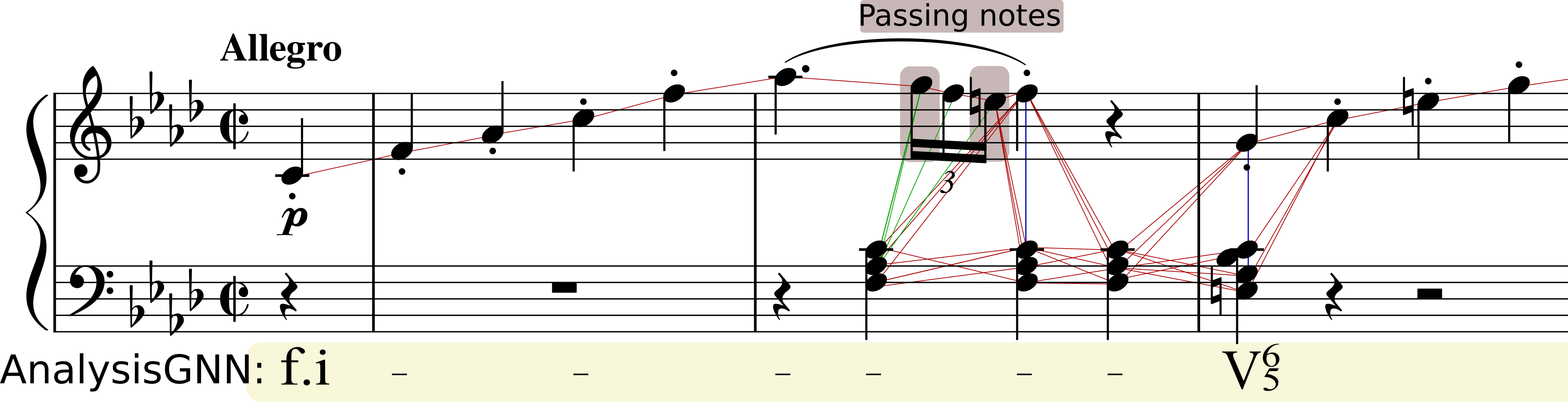}
    \caption{AnalysisGNN predictions on the opening of Beethoven Sonata No.1, Op.2 Movement 1. The figure shows the score graph where red edges connect consecutive notes, blue edges link notes with the same onset, and green edges connect notes with onsets between the onset and offset of another. The model outputs harmonic analysis while highlighting passing notes.}
    \label{fig:prediction-example}
\end{figure}

\subsection{Discussion}

Modern machine learning methods generally benefit from larger amounts of data; however, in music analysis, the lack of a definitive ground truth complicates this, since labels rely on expert judgments to resolve the inherent ambiguity in an over-determined score (e.g., a set of pitch classes can correspond to more than one correct chord label). These subjective interpretations, along with the specific annotation standards and encoding formats employed, heavily influence the resulting labels. Moreover, current evaluation metrics are not equipped to fully capture this ambiguity. In many instances, automatic predictions that deviate from the target labels do not necessarily represent errors but rather alternative, equally compelling interpretations. For example, slight discrepancies in the precise prediction of some tasks may be acceptable, such as a third-relation between chordal roots being more tolerable than a second-relation.

Our efforts into addressing annotational issues in this work were focused on making more informed predictions by the model. When we gain insight on the functional role of every single note in the score, we can better decide and interpret a model's prediction in the music analysis content. Figure~\ref{fig:prediction-example} presents a score example from the beginning of Beethoven Sonata in F minor, Op.~2/1, illustrating how AnalysisGNN not only identifies the underlying harmony but also highlights passing notes. In the accompanying score graph, which is what AnalysisGNN uses as input, red edges connect consecutive notes, blue edges link notes with identical onset times, and green edges indicate connections between notes whose onsets occur between the onset and offset of another note.

By extending this insight to our other tasks, i.e. cadence detection, phrase and section boundary identification, and metrical position estimation, we observe a compounded benefit. Knowing which notes are functionally relevant sharpens the model’s ability to pinpoint cadential patterns, delineate phrase structures, and correctly assign metrical strengths. For instance, filtering out non-chord-tones reduces false positives in cadence boundaries, while emphasizing chord-tones helps align phrase onsets with structurally important events. Likewise, accurate note‐level functionality supports section segmentation by reinforcing thematic or harmonic anchors. Taken together, the effects of all combined analysis task underscore the value of a unified analysis perspective. We trust these improvements in note‐level filtering and labeling could also propagate upward to higher‐level or reductive analyses, resulting to more coherent and musically meaningful predictions.

\section{Conclusion}

In this paper, we presented \emph{AnalysisGNN}, a unified graph neural network framework for symbolic music analysis with over 20 individual analysis tasks with note-level granularity. By integrating a dedicated passing note prediction we achieve cleaner label signals and faster inference without sacrificing training stability. This design allows AnalysisGNN to leverage heterogeneous corpora and annotation schemes, yielding competitive performance across cadence detection, Roman numeral analysis, phrase and section identification, and metrical position estimation, while demonstrating strong resilience to domain shifts and annotation inconsistencies.

Our experimental results demonstrate that \emph{AnalysisGNN} achieves comparative accuracy to traditional single-task models for some tasks but most importantly it exhibits resilience to domain shifts and annotation inconsistencies. Inspired by recent advances such as RNBert, we plan to explore self-supervised pretraining of the GNN encoder to further boost performance. Moreover, recognizing that musical analysis is inherently ambiguous, where multiple interpretations can be equally valid, we advocate for the development of new evaluation metrics that move beyond binary right/wrong judgments. We believe a new evaluation standard will bring automatic analysis closer to its inherent music-theoretical contingencies and do better justice to the individual tasks at hand.  

\section*{Acknowledgements}
This work is supported by the European Research Council (ERC) under the EU's Horizon 2020 research \& innovation programme, grant agreement No.\ 101019375 (\textit{Whither Music?}).

\bibliographystyle{splncs04}
\bibliography{biblio.bib}
\end{document}